# Collective properties of deformed atomic clusters described within a projected spherical basis

A. A. Raduta, Al. H. Raduta and R. Budaca

Institute of Physics and Nuclear Engineering, Bucharest, POB MG6, Romania

Abstract. Several relevant properties of the Na clusters were studied by using a projected spherical single particle states. The proposed model is able to describe in an unified fashion the spherical and deformed clusters. Photoabsorbtion cross section is realistically explained within an RPA approach and a Shiff dipole moment as a transition operator.

Key words: atomic clusters, cluster shape, polarizability, plasmon energy, photoabsorbtion, Shiff-dipole

## 1. INTRODUCTION

Formalisms used to describe atomic clusters depend essentially on their size. Thus for clusters having only few constituents, the ab initio quantum chemical methods [1] are vastly used. Large clusters are most comfortably described by statistical models or semiclassical approaches [2]. Even for medium sized clusters, statistical methods offer a suitable frame for a qualitative interpretation. For such situations all adopted formalisms are based on the mean field approach which allows to replace a many body composite system consisting of valence electrons interacting mutually (which can be easily delocalized or excited by pumping an energy in optical domain) and with the correlated ionic core by a system of interacting electrons moving in an average potential. In this way the static properties of metallic clusters are quantitatively described by the substitute system Hamiltonian. Models are distinguished by the specific derivation of the mean field. Three solutions for the average potential are to be mentioned: i) solving the Kohn-Sham equations [3]. ii) assuming the positive charge of the ionic core uniformly distributed in a sphere of radius R defining the cluster size. This is known as the jellium hypothesis [4] iii) postulating the expression of the average potential. Several reviews, on the fundamental features of metallic clusters, have been published [6].

Since the shell structure and magic clusters are associated with a spherical symmetry, most of theoretical investigations regard spherical clusters. The first paper devoted to the deformed clusters appeared in 1985 [7]. Several details of the mass spectrum [8], not described by spherical jellium models, are clearly explained. Also the fragmentation of collective states can be consistently described [9]. The formalism is known as Clemenger-Nilsson (CN) model since the mean field is similar to that used by Nilsson model with the difference that in the new version the spin-orbit term is missing.

Although the model was very successful in describing many properties of deformed clusters there are some limitations due to the fact that the rotational symmetries are broken. Several years ago we proposed a new schematic model based on a projected spherical single particle basis [10]. The main properties of the single particle states which make them very useful are:1) are mutually orthogonal 2) depend on a parameter simulating the quadrupole deformation  3) in the spherical limit the set should be identical to the spherical shell model, while for nonvanishing deformation the associated energies are identical to those produced by the CN model. Moreover, the projected states approximate quite well the projected Clemenger Hamiltonian eigenstates. Once such a basis is constructed, the RPA formalism for spherical and deformed clusters can be unitarily applied. The path to this goal is organized according to the following plan. The projected spherical basis is constructed in Section 2. Several testing applications are briefly presented. The dipole excitations are described within the random phase approximation (RPA) in Section 3, while the final conclusions are drawn in Section 4.

## 2. THE PROJECTED SPHERICAL SINGLE PARTICLE BASIS

The single particle basis is defined by using a particle-core Hamiltonian

$$H = \frac{p^2}{2m} + \frac{m\omega_0^2 r^2}{2} + D\left(l^2 - <l^2>\right) + H_c - m\omega_0^2 r^2 \sum_{\lambda=0,2} \sum_{-\lambda \leq \mu \leq \lambda} \alpha^*_{\lambda\mu} Y_{\lambda\mu}, \qquad (2.1)$$

where $\alpha$ are the $\lambda$–pole coordinates of the surface of a phenomenological core. The monopole coordinate is fixed by using the volume conservation condition during the deformation process. The quadrupole coordinates defines the quadrupole bosons by:

$$\alpha_{2\mu} = \frac{1}{k\sqrt{2}}\left[b^\dagger_{2\mu} + (-)^\mu b_{2-\mu}\right]. \qquad (2.2)$$

Consider now the coherent state:

$$|\Psi_c> = \exp\left[d\left(b^\dagger_{20} - b_{20}\right)\right]|0>, \qquad (2.3)$$

with d a deformation parameter which simulates the quadrupole deformation of the core. Note that averaging H with the coherent state one obtains the CN Hamiltonian provided the deformations specific to the two schemes are related by *0.693kδ=d*. If the core Hamiltonian is harmonic in the quadrupole bosons, then averaging H on the shell model state | NlI > one obtains a quadrupole boson Hamiltonian which admits the coherent state as an eigenstate. These two features allow us to approximate the eigenvalues of H by the average values corresponding to the projected states:

$$\phi_{IM;\sigma}(nl;d) = N_{nl}^I(d)\left[P_{MI}^I \mid nlI > \Psi_c(d)\right]\chi_\sigma, \text{ for } I \neq 0, l = \text{even},$$
$$\phi_{00;\sigma}(nl;d) = N_{nl}^0(d)\left[P_{00}^0\left[\mid nl > \hat{s}\right]_{l+1,0} \Psi_c(d)\right]\chi_\sigma, \text{ for } I = 0, l = \text{odd},$$
(2.4)

where $N_{nl}^I(d)$ denotes the normalization factor [13], which is given in Appendix A, while $P_{MK}^I$ the projection operator. The result for single particle energies is:

$$\epsilon_{nl}^I(d) = \hbar\omega_0\left(N+\frac{3}{2}\right) - D\left[l(l+1) - \frac{N(N+3)}{2}\right] + \hbar\omega_0\left(N+\frac{3}{2}\right)\frac{1}{90}(\Omega_\perp^2 - \Omega_z^2)^2\left[1 + \frac{1}{d^2}\left\langle\sum_\mu b_{2\mu}^\dagger b_{2\mu}\right\rangle\right]$$
$$-\hbar\omega_0\left(N+\frac{3}{2}\right)\frac{1}{3}(\Omega_\perp^2 - \Omega_z^2)F_{I1}$$
(2.5)

where $\Omega_z$ and $\Omega_\perp$ are the frequencies corresponding to the mentioned directions used by the CN model, while $F_{I1}$ are geometrical factors. The parameters involved in the expression of s.p. energies are: $D = -0.04\hbar\omega_0$, $\hbar\omega_0 = E_F N^{-1/3}$, $E_F = 3$ eV.

The s.p. energies and projected wave functions have been used to calculate the equilibrium deformations, the second difference for the total energy and the electron density. The total energy for a N-cluster is obtained by summing the single particle energies, in an increasing order, at a given deformation parameter d. The equilibrium deformation is that value of d which makes the total energy minimum. The second difference for the total energy is defined by:

$$\Delta_2(N) = \frac{3}{4}\left\{\left[E(N+1) - E(N)\right] - \left[E(N) - E(N-1)\right]\right\},$$
(2.6)

while the electron density is:

$$\rho = \sum \frac{\nu(I)}{2I+1}\mid\Phi_{IM,1/2}(n,l)\mid^2, \quad \nu(I) = 4 - 2\delta_{I0},$$
(2.7)

where the summation is performed over all occupied states.

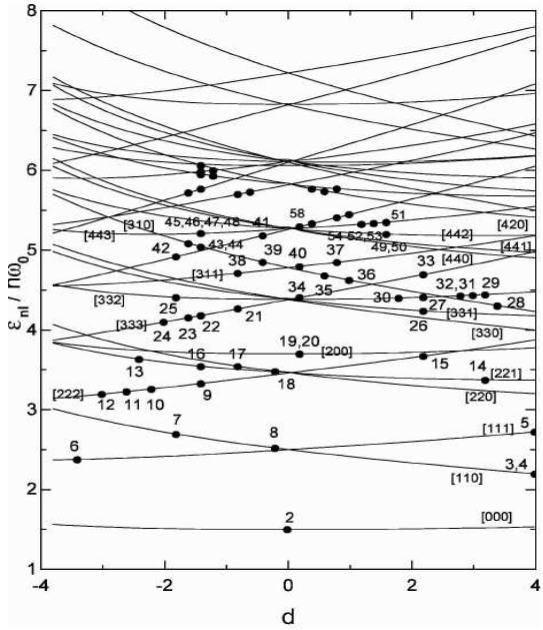
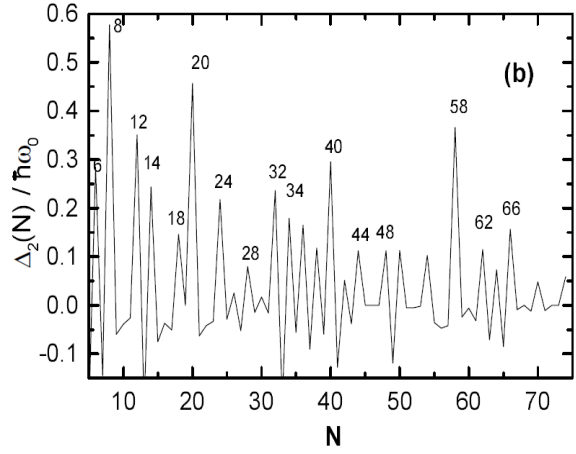

Fig. 1. s.p. energies as function of the deformation d.

Fig.2. Second difference for the total energy as function of N.

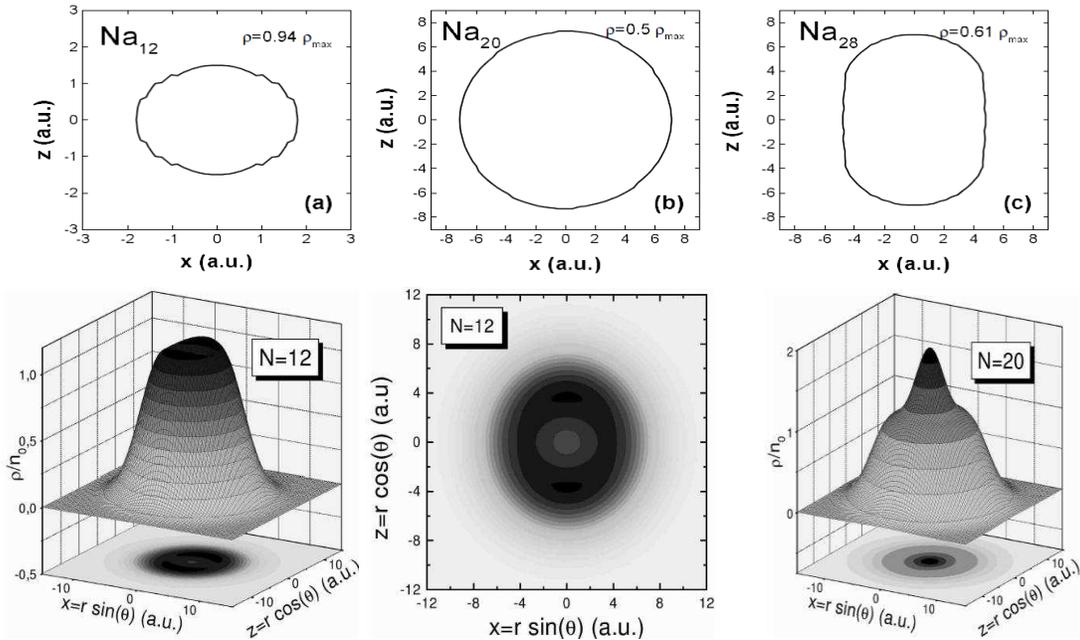

Fig. 3. In the first row we present the curves characterized by a constant electron density. In the second row the spatial distribution for density is presented. For the cluster $Na_{12}$ the projection in the xz plane is also given.

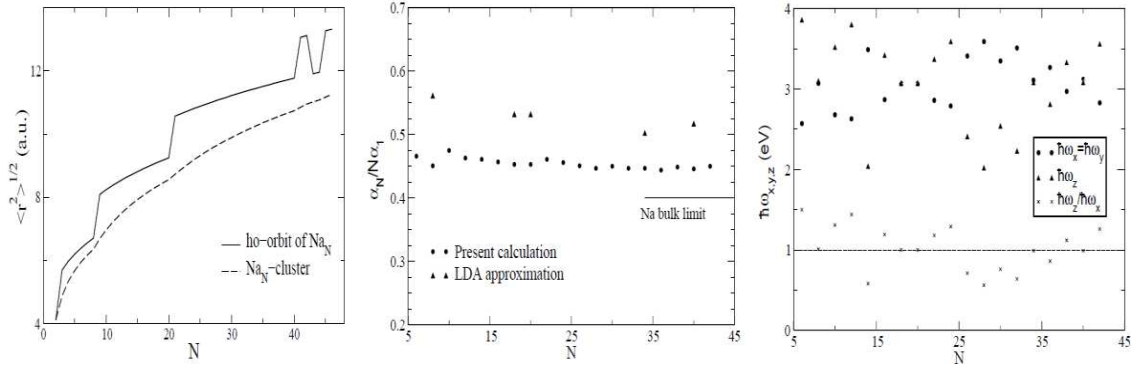

| Fig4.r.m.s. values for h.o. state as function of N | Normalized polarizability as function of N. | Plasmon frequencies as function of N |

A measure for the spatial extension of a cluster is the r.m.s. value associated to the last atom in the cluster, i.e. the highest occupied state:

$$\langle r^2 \rangle_{ho}^{1/2} = \left[ \int r^2 \, | \Phi_{IM;1/2}(nl;d) |^2 \, d^3r \right]^{1/2}. \qquad (2.8)$$

The electric polizability is calculated in terms of the number of spilled out electrons:

$$\alpha = R^3 \left( 1 + \frac{N_{sp}}{N} \right), \quad N_{sp} = \int_{r>R} \rho \, d^2r, \qquad (2.9)$$

while the plasmon energy along the i(=x,y,z) direction is determined by the number of electrons inside the jellium sphere and the time derivatives of the background potential [10]:

$$\hbar \omega_i = \hbar \omega_{Mie} \left[ f_i \frac{N_{ins}}{N} \right]^{1/2} \text{[a.u.]}, \; i = x, y, z, \quad f_i = \frac{1}{\omega_0^2} \frac{\partial^2 V_{bg}}{\partial x_i^2}. \qquad (2.10)$$

$\omega_{Mie}$ denotes the Mie frequency $\hbar \omega_{Mie} = r_s^{-3/2} \text{[a.u]}$, with $r_s$ being the Wigner-Seitz radius which for Na clusters is about 3.93 a.u. The results for r.m.s. of the highest occupied state, polarizability and plasmon frequency are given in Fig.4 as function of the number of components [11]. The left panel of Fig.4 suggests that $Na_{41}$ exhibits a halo structure.

## 3. THE RPA DESCRIPTION OF THE DIPOLE COLLECTIVE STATES

Throughout this paper applications refer to the alkali clusters of Na. The valence electrons are assumed to move in the mean field, presented in the previous chapter, and interact among themselves through the Coulomb interaction. This interaction is expanded in multipoles from which we retain only the dipole term. The resulting many body Hamiltonian, consisting of the mean field one body term and a separable dipole-dipole interaction is treated by the RPA formalism [12]. The amplitudes of the phonon operator

$$C^{\dagger}_{1\mu} = \sum_{ph} \left[ X^n_{ph}(c^{\dagger}_p c_h)_{1\mu} - Y^n_{ph}(c^{\dagger}_h c_p)_{1\mu} \right], \quad (3.1)$$

where obtained by solving the RPA equations. The excitation energies $\omega_n$ determine the amplitudes $X^n$, $Y^n$. The reduced probability for exciting the system from the ground state $|0^+\rangle$ to the RPA state $|1^-_n\rangle$ is given by:

$$B(E1, 0^+ \rightarrow 1^-_n) = |\langle 0^+ \| M(E1) \| 1^-_n \rangle|^2, \quad (3.2)$$

where the transition operator is taken to be a Shiff-like dipole operator:

$$M(E1) = \sqrt{\frac{4\pi}{3}} e Y_{1\mu}(\Omega) \left( r - \frac{3}{5} \frac{r^3}{r_s^2} \right). \quad (3.3)$$

The correction to the dipole operator is needed to simulate the screening effect of the particle-core interaction due to the electronic cloud. In our model the cubic term is responsible for the description of the volume type excitations. The photoabsorbtion cross section is obtained by folding the individual transitions by a Lorentzian characterized by a damping factor $\gamma = \Gamma/\omega_r$ ranging from 0.06 to 0.135:

$$\sigma(\omega) = C \sum_n f_n L(\omega; \omega_n, \Gamma_n). \quad (3.4)$$

Here $f_n$ denotes the oscillator strength and C a normalization factor which is equal to 1.0975 (eV Å$^2$). Comparison with the experimental data is made in Fig. 5 in terms of the photoabsorbtion cross section for clusters Na with various number components.

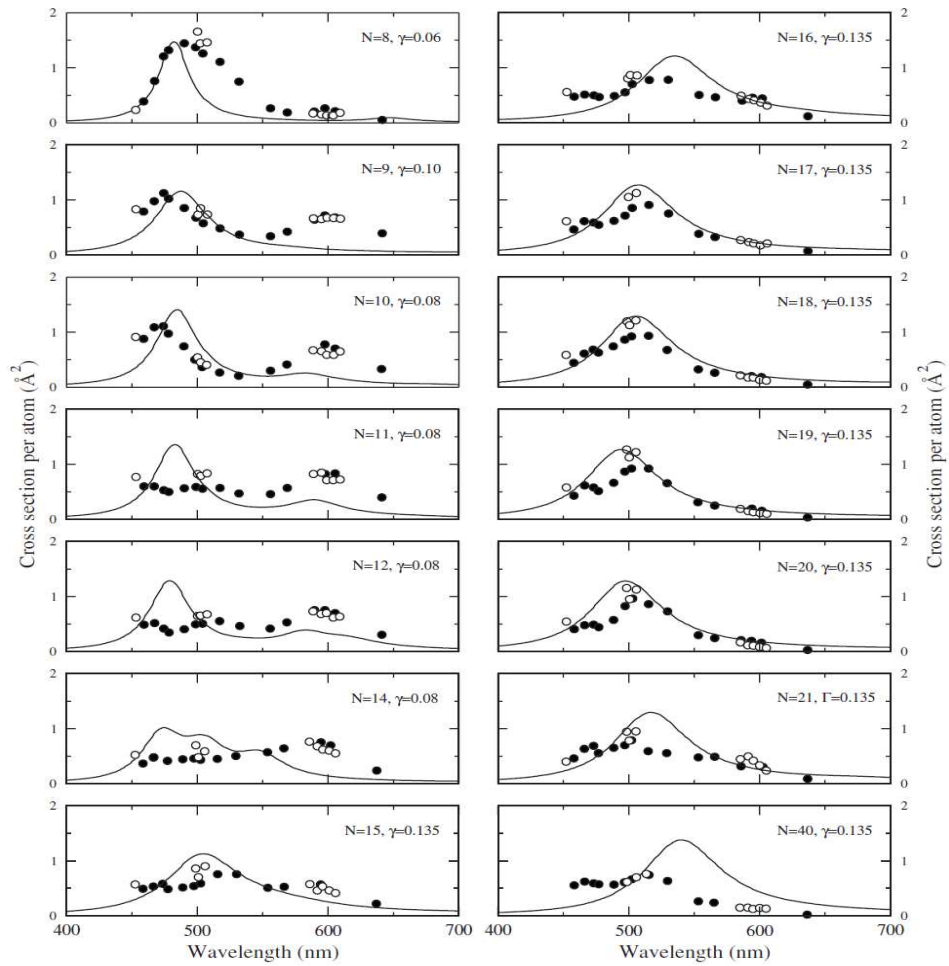

Fig. 5. Photoabsorbtion cross section per atom as function of excitation wavelength.

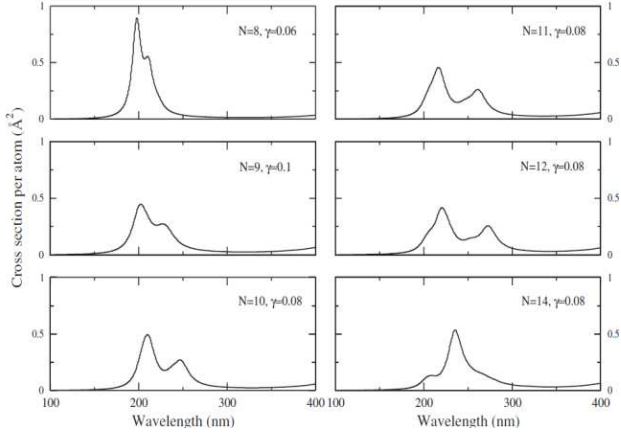 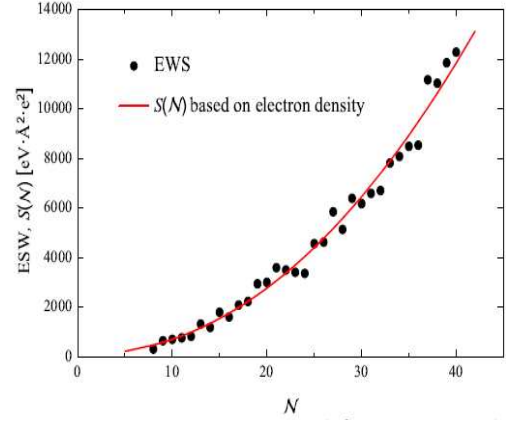

Fig. 6. Photoabsorbtion cross section for the volume like excitations.

Fig. 7. EWS and $S(N)$ as function of the components number.

Taking into account the $N$ dependence of the RPA energies one may say that the states shown in Fig. 5 describe surface type of plasmon oscillations while those from Fig. 6 are of volume type. Both kinds of excitations are fragmented due to the quadrupole deformation of the mean field. Taking for the electron density a Fermi type expression one finds out that the Shiff dipole moment satisfies a sum rule [12]:

$$\text{EWS} \equiv \sum_n (E_n - E_0) |\langle 0 \| M(E1) \| 1_n \rangle|^2$$
$$= \frac{9\hbar^2 e^2}{2m} \left[ N - \frac{6}{5} N^{2/3} + \frac{99}{175} N^{4/3} + \frac{\pi^2 a^2}{5 r_s^2} \left( \frac{594}{35} N^{2/3} - 14 \right) \right] \equiv S(N), \quad (3.5)$$

with $a(N) = -0.975157 - 0.0112138 N^{1/3} + 0.360518 N^{2/3}$.

In Fig. 7 one shows that indeed, the equality $\text{EWS} = S(N)$ holds.

Using the RPA states, the number of the spilled out electron is calculated and then the electric polarizability for the $N < 40$ clusters are calculated. The agreement with experimental data is very good.

## 4. CONCLUSIONS

The results presented in the previous sections prove that the projected spherical single particle basis is suitable for a realistic and an unitary description of the collective properties of atomic clusters of spherical and deformed shapes. A new sum rule associated with the Shiff dipole moment is pointed out.

## 5. APENDIX A

The norms of the angular momentum projected states from Eq. (2.4) have the expression:

$$[N_{nl}^I]^{-2} = \sum_J (C_{I0I}^{lJI})^2 (N_j^{(c)})^{-2}, \text{ for } I \neq 0, l = even,$$

$$[N_{nl}^0]^{-2} = \frac{1}{4} \frac{1}{2l+3} (N_{l+1}^{(c)})^{-2}, \text{ for } I = 0, l = odd, \quad (A.1)$$

With the standard notations fot the Clebsch Gordan coefficients and $N_J^{(c)}$ denoting the norm of the J component projected from the deformed state describing the core. This norm depends on the deformation parameter d and was calculated in Ref. [13]. The analytical expression obtained therein is:

$$N_J^{(c)}(d) = \left[(2J+1)I_J^{(0)}(d)\right]^{-1/2} e^{d^2/2},$$

$$I_J^{(0)}(d) = \frac{(J!)^2}{(\frac{1}{2}J)!(2J+1)!}(6d^2)^{J/2} e^{-d^2/2} F(\frac{1}{2}(J+1), J+\frac{3}{2}; \frac{3}{2}d^2). \quad (A.2)$$

Here the notation F(a,b;x) stands for the degenerate hypergeometrical function.

# References

1. V. Bonacic-Koutecky, P. Fantuccci and J. Koutecky, Phys. Rev. B37, 4369 (1988).

2. M. Cini, J. Catal., 37, 187 (1975).

3. W. Kohn and L. J. Sham, Phys. Rev. A140, 1133 (1965).

4. J. L. Martins, J. Buttet and R. Cae, Phys. Rev. B31, 1884 (1985).

5. D. E. Beck, Solid State Communic. 49, 381 (1984).

6. H. Nishioka, K. Hansen and B. Mottelson, Phys. Rev. B42, 9377 (1990).

7. W. D. Knight, et al., Phys. Rev. Lett., 52, 2141 (1984).

8. M. Y. Chou and M. L. Cohen, Solid State Commun. 52, 645 (1984).

9. A. A. Raduta, Ad. R. Raduta, al. H. Raduta, Phys. Rev. B 59, 8209 (1999).

10. A. A. Raduta, E. Garrido, E. Moya de Guerra, Eur. Phys. J. D 15, 6577 (2001).

11. A. A. Raduta, R. Budaca, Al. H. Raduta, Phys Rev. A 79, 023202 (2009).

12. A. A. Raduta, R. Budaca, to be published.

13. A. A. Raduta, et al., Nucl. Phys. A 381, 253-276 (1982)